\documentstyle[prl,twocolumn,aps]{revtex}
\input{epsf}
\begin{document}
\draft
\twocolumn[\hsize\textwidth\columnwidth\hsize\csname @twocolumnfalse\endcsname
\title{Emergence of Quantum Chaos in Quantum Computer Core and How to Manage It}

\author{B. Georgeot and D. L. Shepelyansky$^{(*)}$}

\address {Laboratoire de Physique Quantique, UMR 5626 du CNRS, 
Universit\'e Paul Sabatier, F-31062 Toulouse Cedex 4, France}

\date{May 2, 2000}

\maketitle

\begin{abstract}
We study the standard generic quantum computer model, 
which describes a realistic isolated quantum computer with fluctuations
in individual qubit energies and residual short-range inter-qubit couplings.
It is shown that in the limit where the  fluctuations and couplings
are small compared to one-qubit energy spacing the
spectrum has a band structure and a renormalized Hamiltonian
is obtained which describes the eigenstate properties
inside one band. The studies are concentrated on the central
band  of the computer (``core'') with the highest density
of states. We show that
above a critical inter-qubit coupling strength, quantum chaos sets in, 
leading to quantum ergodicity 
of the computer eigenstates. In this regime the ideal
qubit structure disappears, the eigenstates become complex and the operability
of the computer is quickly destroyed. We confirm that
the quantum chaos border decreases only linearly with the number of qubits
$n$, although the spacing between
multi-qubit states drops exponentially with $n$.  The investigation
of time-evolution
in the quantum computer shows that in the quantum chaos regime,
an ideal (noninteracting) state quickly disappears
and exponentially many  states
become mixed after a short chaotic time scale
for which the dependence on system parameters is determined. 
Below the quantum chaos border
an ideal state
 can survive for long times and be used for computation.
The results 
show that a broad parameter region does exist where the efficient
operation of a quantum computer is possible.
\end{abstract}
\vskip1pc]

\narrowtext
\section{Introduction}
During the last decade, a remarkable progress has been achieved in
the fundamental understanding of the main elements necessary for
the creation of a quantum computer.   Indeed, as stressed  by Feynman
\cite{feynman}, classical computers have tremendous problems to simulate
very common quantum systems, since the computation time grows exponentially
with the number of quantum particles.  Therefore for such problems
it is natural to envision a computer composed from quantum elements (qubits)
which operate according to the laws of quantum mechanics.  In any 
case, such devices will be in a sense unavoidable since the technological
progress leads to chips of smaller and smaller size which will eventually
reach the quantum scale.  At present a quantum computer is viewed as a
system of $n$ qubits (two-level quantum systems), with the possibility
of switching on and off a coupling between them (see the detailed reviews
in \cite{DiVi1,josza,steane2}).
The operation of such computers is based
on reversible unitary transformations in the Hilbert space whose dimension
$N_H = 2^n$ is exponentially large in $n$.  It has been shown that all
unitary operations can be realized with two-qubit transformations
\cite{deutsch,DiVi2}.
This makes necessary the existence of a coupling between qubits. 
Any quantum algorithm will be a sequence of such fundamental
transformations, which form the basis of a new quantum logic.

An important next step was the discovery of quantum algorithms which can 
make certain computations
much faster than on a classical computer.  The most impressive is the problem
of factorization of large numbers in prime factors, for which Shor
constructed  \cite{shor1} a quantum algorithm which is exponentially faster than 
the classical ones.  It was also shown by Grover \cite{Grover} that
the searching of an item in a long list is parametrically
much faster on a quantum computer.
The recent development of error-correcting codes \cite{shor2,steane1} showed
that a certain amount of noise due to external coupling could be tolerable 
in the operation of a quantum computer. 

All these exciting developments motivated a great body of experimental
proposals to effectively
realize such a quantum computer.  
They include ion traps \cite{zoller}, nuclear magnetic resonance systems
\cite{nmr}, nuclear spins with interaction controlled electronically
\cite{vagner,kane} or by laser pulses \cite{bowden}, quantum dots \cite{loss}, Cooper pair boxes
\cite{cooper}, optical lattices \cite{lattice} and electrons floating on liquid 
helium \cite{helium}.  As a result, a two-qubit gate has been experimentally
realized with cold ions \cite{monroe}, and the Grover algorithm
has been performed for three qubits made from nuclear spins in a molecule 
\cite{3q}.  However, to have a quantum computer competitive with a classical
one will require a much larger number of qubits.
For example, the minimal number of qubits for which Shor's algorithm
will become useful is of the order of $n=1000$ \cite{steane2}.
As a result, a great
experimental effort is still needed on the way to quantum computer realization.

A serious obstacle to the physical realization of such computers is 
the quantum decoherence due to the couplings with the external world
which gives a finite lifetime to the excited state of a given qubit.
This question has been discussed by several groups for different
experimental qubit realizations \cite{steane2,DiVi2,haroche,vagner2}.
The effects of decoherence and laser pulse shape broadening were 
numerically simulated in the context of Shor's algorithm \cite{paz,zurek},
and shown to be quite important for the operability of the computer.
However,  in a number of physical proposals, for example nuclear
spins in two-dimensional semiconductor structures, the relaxation time
due to this decoherence process can be many orders of magnitude larger
than the time required for the gates operation 
\cite{DiVi1,vagner,kane,vagner2},
so that there are hopes to manage this obstacle.

Here we will focus on a different obstacle to the physical realization
of quantum computers that was not stressed up to now.  This problem
arises even if the decoherence time is infinite and the system is 
isolated/decoupled 
from the external world.  Indeed, even in the absence of decoherence there
are always imperfections in physical systems.  Due to that the spacing
between the two states of each qubit will fluctuate in some finite 
detuning interval $\delta$.  Also, some residual static interaction $J$ between
qubits will be unavoidably present (we remind that an inter-qubit coupling 
is required to operate the gates).  Extensive studies of many-body interacting
systems such as nuclei, complex atoms,
quantum dots and quantum spin glasses
\cite{french,aberg,zelevinsky,sivan,1997,jacquod,mirlin,geor1,pichard,georgeot}
have shown that generically
in such systems the interaction leads to quantum chaos characterized
by ergodicity of the eigenstates and level spacing statistics 
as in  Random Matrix Theory (RMT) \cite{houches,guhr}.  In a sense
the interaction leads to dynamical thermalization without coupling to
an external thermal bath.  If the quantum computer were in such a regime,
its operability would be effectively destroyed since the noninteracting
multi-qubit states representing the quantum register states will
be eliminated by quantum ergodicity.  
 
In this respect, it is important to stress that unavoidably the residual 
interaction $J$
will be much larger than the energy spacing $\Delta_n$ 
between adjacent eigenstates
of the quantum computer.   Indeed the residual interaction $J$ 
is relatively small so that all $N_H$ computer eigenenergies are
distributed in an energy band of size $\Delta E \sim n \Delta_0$, where
$\Delta_0$ is the average energy distance between the two levels of one
qubit and $n$ is the total number of qubits in the computer.  As a consequence,
the spacing between multi-qubit states is $\Delta_n \approx \Delta E /N_H
\sim n\Delta_0 2^{-n} \ll \Delta_0$.  Let us consider a realistic estimate
for $\Delta_n$ and $J$ for the case with $n=1000$ as required for Shor's
algorithm to be useful.  For $\Delta_0 \sim 1$ K, which corresponds to the 
typical one-qubit spacing in the experimental proposals \cite{vagner,kane},
the multi-qubit spacing becomes $\Delta_n \sim 
10^3 \times 2^{-10^3} \Delta_0 \sim 10^{-298}$ K.  This value will definitely
be much smaller than any physical residual interaction.  In the case of
the proposal \cite{kane}, for example, with a distance
between donors of $r=200 $ {\AA} and an effective 
Bohr radius of $a_B=30$ {\AA} ( Eq.(2) of \cite{kane}), the coupling between qubits
(spin-spin interaction)
is $J \sim \Delta_0 \sim 1$ K. By changing the electrostatic gate potential,
the effective electron mass can be modified up to a factor of two.
Since $J  \propto (r/a_B)^{5/2} 
\exp (-2r/a_B)/a_B$, and $a_B$ is inversely proportional to the effective mass,
this gives a minimal residual spin-spin 
interaction of $J \sim 10^{-5}$ K $\gg \Delta_n$.   In this situation, one
would naturally/naively expect that level mixing, quantum ergodicity
of eigenstates and chaos are unavoidable since the interaction is much bigger
than the energy spacing between adjacent levels ($J \gg \Delta_n$).

In spite of this natural expectation, it was shown recently in \cite{GS}
that in the quantum computer 
the quantum chaos sets in only for couplings $J$ exponentially stronger than 
$\Delta_n$.  In fact, it was shown that the critical coupling $J_c$ for
the transition to quantum chaos decreases only linearly with the number
of qubits $n$ (for short-range inter-qubit coupling): $J_c \sim \Delta_0/n$.
This result opens a broad parameter region where a quantum computer can be
operated below the quantum chaos border, when noninteracting multi-qubit
states are very close to the exact quantum computer eigenstates.
For example, at $n=1000$ and $\Delta_0 \sim 1$ K, the critical residual
interaction is $J_c \sim 1$ mK, compatible with the proposal discussed above
\cite{kane}.  

In the present paper, we study in more details the transition to chaos
and how it affects the time evolution of the system.  The effects of
residual interaction in the presence or absence of
fine fluctuations of individual qubit energy spacing are 
analyzed in great detail.  The paper is composed as follows.  In the
next section we describe the standard generic quantum computer (SGQC) model, 
introduced in \cite{GS}.  In section III, we present the result of
numerical and analytical studies of eigenenergies and eigenstate properties 
of this model.   Section IV is devoted to the analysis of the time evolution
of this system, and the typical time scales for the development of quantum
chaos are presented as a function of the system parameters.  We end by some 
concluding remarks in the last section.

\section{Standard generic quantum computer model}

In \cite{GS} the standard generic quantum computer (SGQC) model was introduced
to describe a system of $n$ qubits containing imperfections which generate
a residual inter-qubit coupling and fluctuations in the
energy spacings between the two states of one qubit.
The Hamiltonian of this model reads:
\begin{equation}
\label{hamil}
H = \sum_{i} \Gamma_i \sigma_{i}^z + \sum_{i<j} J_{ij} 
\sigma_{i}^x \sigma_{j}^x,
\end{equation}
where the $\sigma_{i}$ are the Pauli matrices for the qubit $i$ and the second
sum runs
over nearest-neighbor qubit pairs on a two-dimensional lattice
with periodic boundary conditions applied.
The energy spacing between the two states of a qubit is represented 
by $\Gamma_i$ randomly and uniformly distributed in the interval 
$[\Delta_0 -\delta /2, \Delta_0 + \delta /2 ]$.  The detuning
parameter $\delta$
gives the width of the distribution near the average value $\Delta_0$ 
and may vary from $0$ to $\Delta_0$.  Fluctuations
in the values of $\Gamma_i$ appear generally as a result of imperfections.
For example, in the frame of the experimental proposals \cite{vagner,kane}, 
the detuning $\delta$ will appear for nuclear spin levels as a result
of local magnetic fields and density fluctuations. 
For electrons floating on liquid helium \cite{helium}, it will
appear due to 
fluctuations of the electric field near the surface. 
The couplings $J_{ij}$ represent the residual static interaction 
between qubits which
is always present for reasons explained in the introduction.
They can originate from spin-exciton exchange \cite{vagner,kane},
Coulomb interaction \cite{zoller}, 
dipole-dipole interaction \cite{helium}, etc...
To catch the general features of the different proposals,
we chose $J_{ij}$ randomly and uniformly distributed in the interval $[-J,J]$. 
We note that a similar Hamiltonian, but without coupling/detuning
fluctuations,
 was discussed for a quantum computer based
on optical lattices \cite{lattice,molmer}.  This SGQC model describes
the quantum computer hardware, while the gate operation in time should
include additional time-dependent terms in the Hamiltonian (\ref{hamil})
and will be studied separately.  At $J=0$ the noninteracting
eigenstates of the
SGQC model  can be presented as $|\psi_i>
=|\alpha_1,...,\alpha_n>$ where $\alpha_k=0,1$ marks the polarization
of each individual qubit.  These are the ideal eigenstates of a quantum
computer, and we will call them quantum register states.  For $J \neq 0$,
these states are no longer eigenstates of the
Hamiltonian, and the new eigenstates are now 
linear combinations of different quantum register states.  We will use the term
multi-qubit states to denote the eigenstates of the SGQC model with interaction
but also for the case $J=0$.

\begin{figure}
\epsfxsize=3.4in
\epsfysize=2.6in
\epsffile{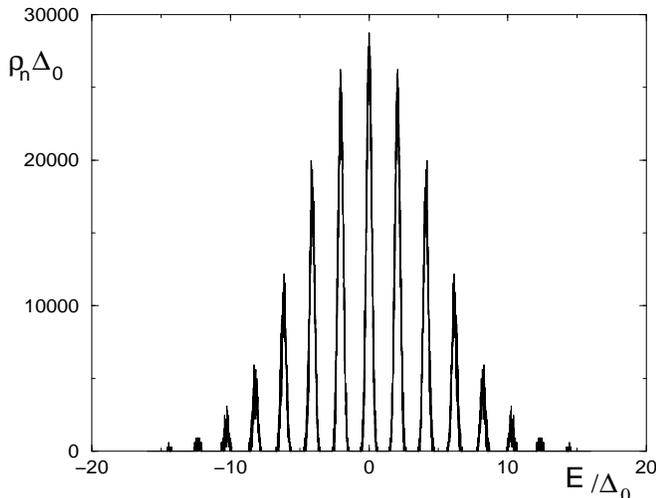}
\caption{Density of multi-qubit states of (\ref{hamil}) 
as a function of total system energy $E$ for $J=0$.  Here 
$n=16$ and $\delta/\Delta_0= 0.2$.  The two extreme bands  at $E/\Delta_0
\approx \pm 16$ contain only
one state and are not seen at this scale.} 
\label{fig1}
\end{figure}

While in \cite{GS} the main studies were concentrated on the case 
where $\delta$ is relatively large and comparable to $\Delta_0$, 
here we will focus on the case $\delta \ll \Delta_0$, which corresponds
to the situation where fluctuations induced by imperfections are relatively
weak.  In this case, the unperturbed energy spectrum of (\ref{hamil}) 
(corresponding
to $J=0$) is composed of $n+1$ well separated bands, with interband spacing
$2\Delta_0$.  An example of the density of multi-qubit
states $\rho_n= 1/\Delta_n$ in this situation is 
presented in Fig.\ref{fig1}.  Since the $\Gamma_i$ randomly fluctuate in
an interval of size $\delta$,  each band at $J=0$, except the extreme ones, have
a Gaussian shape with width $\approx \sqrt{n} \delta $.  The number of states
inside a band is approximately $N_H/n$, so that the energy spacing between
adjacent multi-qubit states inside one band is exponentially small 
($\delta_n \sim n^{3/2} 2^{-n} \delta$), in line with the general
estimate in Section I.
 
In the presence of a residual interaction $J \sim \delta$, the spectrum
will still have the above band structure with exponentially large
 density of states.
For $J \sim \delta \ll \Delta_0$, the interband coupling is
very weak and can be neglected.  In this situation, the SGQC Hamiltonian 
(\ref{hamil}) is to a good approximation described by the 
renormalized 
Hamiltonian $H_{P}=\Sigma_{k=1}^{n+1} \hat{P_k} H \hat{P_k}$ where $\hat{P_k}$
is the projector on the $k^{th}$ band, so that qubits are coupled only
inside one band.  We will thereafter concentrate our studies on the band
nearest to $E=0$. For an even $n$ this band is centered exactly at $E=0$,
while for odd $n$ there are two bands centered at $E=\pm \Delta_0$,
and we will use the one at $E=-\Delta_0$.  Such a band corresponds
to the highest density of states, and in a sense represents 
the quantum computer
core.  It is clear that quantum chaos and ergodicity will first appear 
in this band, which will therefore set
the limit for operability of the quantum computer.  Inside this
band, the system is described by a renormalized Hamiltonian $H_P$ 
which depends only on the number of qubits $n$ and the dimensionless
coupling $J/\delta$.  

\section{Quantum Computer Eigenenergies and Eigenstates}

The first investigations in \cite{GS} showed that the quantum chaos
border in the SGQC model (\ref{hamil})  corresponds to a critical
interaction $J_c$ given by:

\begin{equation}
\label{Jc}
J_c \approx \frac{C \delta}{n},
\end{equation}

where $C$ is a numerical constant.  This border is exponentially
larger than the energy spacing between adjacent multi-qubit states $\Delta_n$.
The physical origin of this difference is due to the fact that
the interaction is of a two-body nature.  As a result, one noninteracting 
multi-qubit
state $|\psi_i>$ has nonzero coupling matrix elements only with $ 2 n$ 
other
multi-qubit states.  In the basis of quantum register states $|\psi_i>$,
the Hamiltonian is represented by a very sparse nondiagonal matrix 
with only $2n+1$ nonzero matrix elements by line of length $N_H=2^n$.
For $\delta \approx \Delta_0$ all these transitions take place 
in an energy interval $B$ of width of order $6 \Delta_0$.  Therefore
the energy spacing between directly coupled states is $\Delta_c \approx B/2n
\approx 3 \Delta_0/n$.  According to the studies of quantum chaos in many-body
systems \cite{aberg,1997,jacquod,mirlin,geor1,pichard,georgeot,GS}, the 
transition to chaos takes place when the matrix elements become larger
than the energy spacing between directly coupled states.  This gives
$J > \Delta_c$ which leads to the relation (\ref{Jc}).  For the case $\delta
\ll \Delta_0$ on which we focus here, still in the renormalized Hamiltonian
$H_P$ the number of nonzero matrix elements in one line is of the order
of $n$, and $B \sim \delta$, so that $\Delta_c \sim \delta /n$, that leads to
the result (\ref{Jc}) \cite{longrange}.

\begin{figure}
\epsfxsize=3.4in
\epsfysize=2.6in
\epsffile{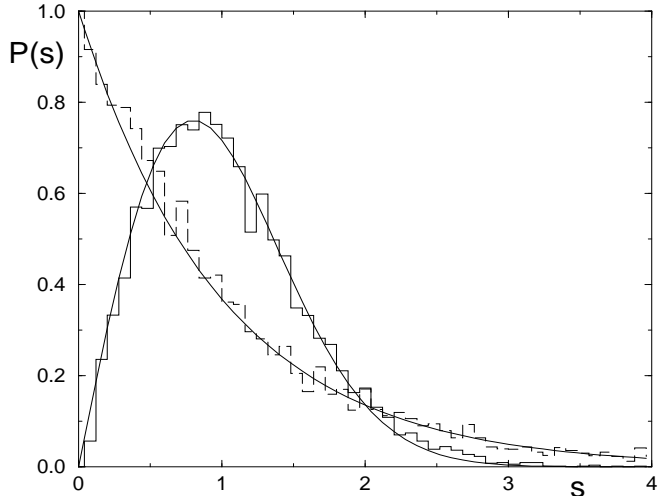}
\caption{Transition from Poisson to WD statistics for the renormalized 
Hamiltonian of SGQC model in the central band.  The statistics is 
obtained for the states
in the middle of the energy band ($\pm 6.25 \% $ around the center) for
$n$=16 : $J/\delta=0.05, \eta=0.99$ (dashed line histogram);
$J/\delta=0.32, \eta=0.047$ (full line histogram). Full curves
 show the Poisson distribution $P_P(s)$ and the Wigner-Dyson distribution
 $P_W (s)$; $N_D=8$, $N_S > 1.2 \times 10^4$.} 
\label{fig2}
\end{figure}

The transition to quantum chaos and ergodicity can be clearly seen in the 
change of the spectral statistics of the system.  One of the most convenient
is the level spacing statistics $P(s)$,  which gives the probability
to find two adjacent levels whose spacing is in $[s,s+ds]$.  Here $s$
is the energy spacing measured in units of average level spacing.  It is well
known that while the average density of states is not sensitive to
the presence or absence of chaos, the fluctuations of the energy spacings
between adjacent levels around the mean value,
determined by $P(s)$, are sensitive to it.
In the presence of chaos, eigenstates are ergodic, 
overlap of wavefunctions gives a finite coupling
matrix element between nearby states and the spectral statistics
$P(s)$ follows the 
Wigner-Dyson (WD) distribution $P_W(s) = (\pi s/2)\exp(-\pi s^2/4)$
typical for random matrices.
This distribution $P_W(s)$ shows level repulsion at small $s$, 
due to the fact that overlap matrix
elements between adjacent levels tend to move them away from each
other.  On the contrary, in the integrable case at $J \ll J_c$, the overlap
coupling matrix element between nonergodic states is very small.
As a result, energy levels are uncorrelated
and $P(s)$ follows the Poisson distribution $P_P(s)=\exp(-s)$ known to be valid
for integrable one-particle systems \cite{houches}.

In the SGQC model, we expect a transition from $P_P(s)$ at small $J$
to $P_W(s)$ above the quantum chaos border (\ref{Jc}).  An example of
such a transition is shown in Fig.2.  To decrease the statistical
fluctuations we averaged over several independent realizations of the 
$\Gamma_i$ and $J_{ij}$ in (\ref{hamil}), which is the standard procedure
used in Random Matrix Theory \cite{houches,guhr}.  We used
up to $N_D=5 \times 10^4$ realizations so that the total statistics
$1.5 \times 10^5 \geq N_S > 1.2 \times 10^4 $. It is interesting to note that in the limit
$J/\delta \rightarrow \infty$ ($\delta \ll J \ll \Delta_0$) the system remains in
the regime of quantum chaos with WD statistics \cite{note}, 
as is illustrated in Fig.3.  This means that in the absence of individual qubit
energy fluctuations, the residual coupling alone leads to chaotic eigenstates.

\begin{figure}
\epsfxsize=3.4in
\epsfysize=2.6in
\epsffile{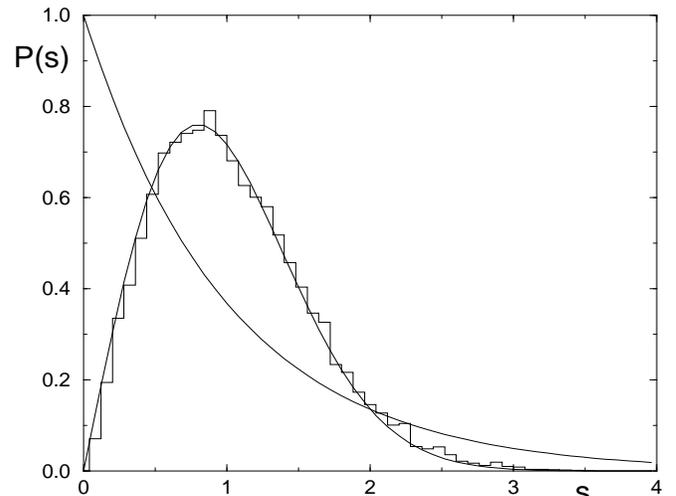}
\caption{Level spacing statistics for the renormalized 
Hamiltonian of SGQC model in the central band for $\delta=0$.  
The statistics is 
obtained for the states
in the middle of the energy band ($\pm 6.25 \% $ around the center) for
$n$=15 : $\eta=0.023$ (histogram).
Full curves
 show $P_P(s)$ and $P_W (s)$; $N_D=20$, $N_S > 1.6 \times 10^4$.} 
\label{fig3}
\end{figure}

To characterize the variation of $P(s)$ from one limiting distribution
to another it is convenient to use the parameter 
$\eta=\int_0^{s_0}
(P(s)-P_{W}(s)) ds / \int_0^{s_0} (P_{P}(s)-P_{W}(s)) ds$ \cite{jacquod},
where  $s_0=0.4729...$ is the intersection point of $P_P(s)$ and $P_{W}(s)$.
In this way $P_P(s)$
corresponds to $\eta=1$, and $P_W(s)$ to $\eta=0$.  The studies of different
systems has already shown that this parameter characterizes well the transition
from one statistics to the other \cite{jacquod,geor1,georgeot,GS}.
Indeed, according to the data of Fig.4, $\eta$ changes from $1$ at small
$J$ to $\eta \approx 0$ at large $J$.  To characterize this transition, 
we chose the critical value $J_c$ by the condition $\eta(J_c)=0.3$.  The
dependence of $\eta$ on the rescaled coupling strength $J/J_c$ shows that
the transition becomes sharper and sharper when $n$ increases (Fig.4).

\begin{figure}
\epsfxsize=3.4in
\epsfysize=2.6in
\epsffile{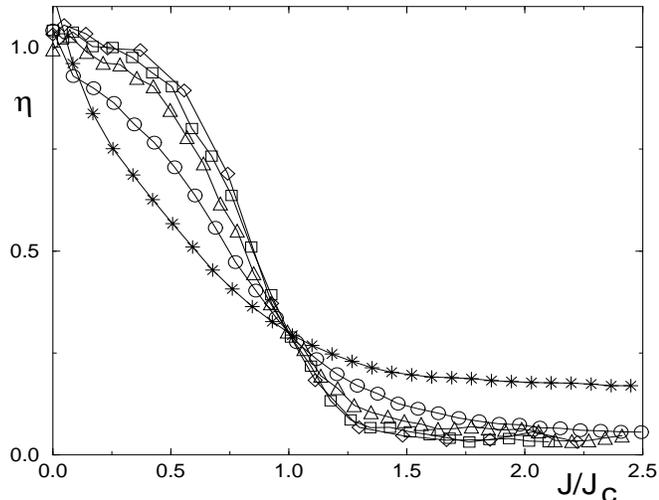}
\caption{Dependence of $\eta$ on the rescaled coupling strength $J/J_{c}$
for the states in the middle of the energy band  for
$n=6 (*),9 $(o)$,12 $(triangles)$,15$(squares)$,16$(diamonds).}
\label{fig4}
\end{figure}

The dependence of the critical coupling strength $J_c$ on the 
number of qubits $n$ is
shown on Fig.5.  It clearly shows that this critical strength decreases
linearly with $n$ and follows the theoretical border (\ref{Jc}) with
$C\approx 3$.  For comparison on the same figure we also show the 
dependence of the multi-qubit spacing $\Delta_n$ (computed numerically)
on $n$.  It definitely demonstrates that $J_c \gg \Delta_n$.

\begin{figure}
\epsfxsize=3.4in
\epsfysize=2.6in
\epsffile{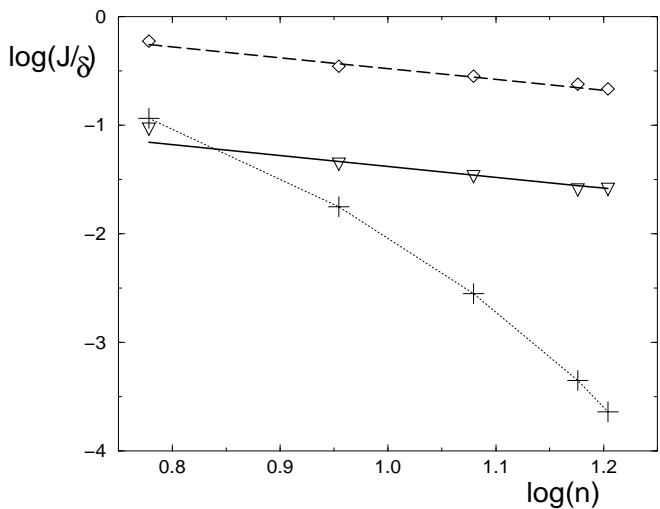}
\vglue 0.2cm
\caption{Dependence of $\log(J_c/\delta)$ (diamonds) 
and $\log(J_{cs}/\delta)$ (triangles) 
versus $\log(n)$; the variation of the scaled multi-qubit
spacing ($\log(\Delta_n/\delta))$ with $\log(n)$ is shown for comparison (+).
Dashed line gives the theoretical formula 
$J_c=C \delta/n$ with $C=3.3$; the solid line is $J_{cs}=0.41 \delta/n$;
the dotted curve is drawn to guide the eye for (+).}
\label{fig5}
\end{figure}

The transition in the level spacing statistics reflects a qualitative
change in the structure of the eigenstates.  While for $J \ll J_c$
the eigenstates are expected to be very close to the quantum register
states $|\psi_i>$, for $J >J_c$ each eigenstate $|\phi_m>$ becomes
a superposition of an exponential number of states $|\psi_i>$.  It is
convenient to characterize the complexity of an eigenstate $|\phi_m>$
by the quantum eigenstate entropy 
$S_q = -\sum_{i} W_{im} \log_2 W_{im}$, where $W_{im}$ 
is the quantum probability
to find the quantum register state $|\psi_i>$ in the eigenstate 
$|\phi_m>$ of the Hamiltonian ($W_{im}=|<\psi_i|\phi_m>|^2$).  
In this way $S_q=0$
if $|\phi_m>$ is one quantum register state ($J=0$), 
$S_q=1$ if $|\phi_m>$ is equally
composed of two $|\psi_i>$, and the maximal value is $S_q=n$ if all
$2^n$ states contribute equally to $|\phi_m>$.  We average $S_q$
over the states in the center of the energy band and $N_D$ realizations
of $\Gamma_i$ and $J_{ij}$.

\begin{figure}
\epsfxsize=3.4in
\epsfysize=2.6in
\epsffile{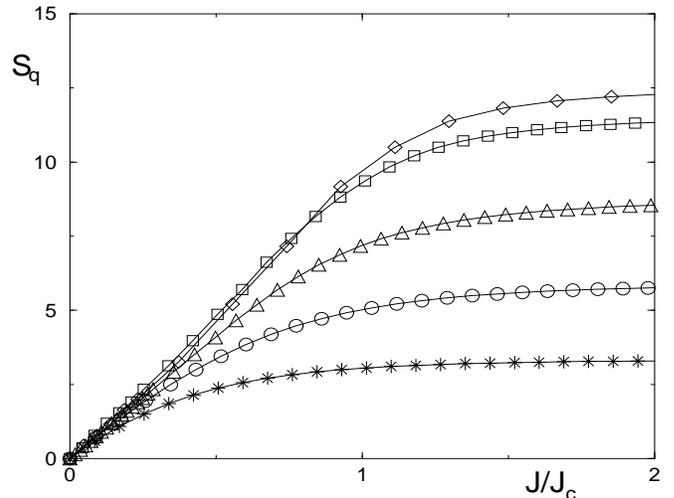}
\caption{Dependence of the quantum eigenstate entropy $S_q$ on $J/J_c$ for
 $n=6 (*)$, 9  (o), 12 (triangles), 15 (squares), 16 (diamonds); 
$1.5 \times 10^5 \geq  N_S > 1.2 \times 10^4$.} 
\label{fig6}
\end{figure}

\begin{figure}
\epsfxsize=3.4in
\epsfysize=2.6in
\epsffile{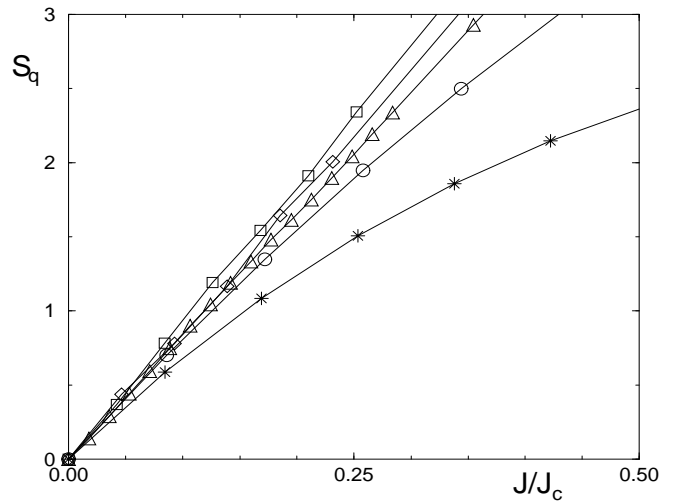}
\caption{Same as Fig.6 but on a larger scale.} 
\label{fig7}
\end{figure}

The variation of this average 
$S_q$ as a function of $J$ for different values of $n$ is shown on Figs.6,7.
It shows that indeed the entropy $S_q$ grows 
with $J$ until it saturates to a large
value corresponding to an exponential number of mixed states.  These data
show that the critical coupling $J_{cs}$ at which $S_q=1$ (two states mixed)
is proportional to $J_c$.  Indeed, Fig.7 shows a small dispersion
near $S_q=1$ when $n$ changes from 6 to 16, while $\Delta_n$ varies by
three orders of magnitude.  This is confirmed by the data on Fig.5,
which give $J_{cs} \approx 0.13 J_c \approx 0.4 \delta/n$.  This result is in
agreement with the results \cite{GS} obtained by direct diagonalization
of the SGQC model (\ref{hamil}) at $\delta \ll \Delta_0$ 
(lower insert in Fig.2 of \cite{GS}).

The quantum eigenstate entropy $S_q$ characterizes the global properties
of the eigenstates, while a more detailed information about them can be
obtained from the local density of states  $\rho_{W}$ 
introduced by Wigner \cite{wigner}:

\begin{equation}
\label{ldos}
\rho_{W}(E-E_{i}) = \sum_{m} W_{im}
\delta(E-E_{m})
\end{equation}

\vskip -0.3cm
\begin{figure}
\epsfxsize=3.4in
\epsfysize=2.6in
\epsffile{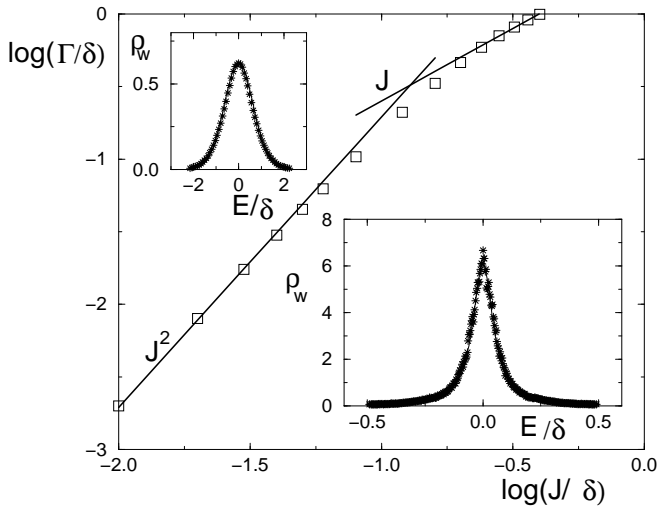}
\vglue 0.3cm
\caption{Dependence of the Breit-Wigner width $\Gamma$ on the
coupling strength $J$ for $n=15$ for the states in the middle of
the energy band.  The straight lines show the theoretical dependence
(\ref{gamma}) with $\Gamma= 1.3 J^2 n/\delta$
and the strong coupling regime with $\Gamma \sim J $; $N_D=20$.
Lower insert: example of the local density of states $\rho_W$
(\ref{ldos}) for $J/\delta=0.08$; the full line shows the best fit of the 
Breit-Wigner form (\ref{bw}) with $\Gamma=0.10 \delta$.  Upper insert:
example of the local density of states $\rho_W$
(\ref{ldos}) for $J/\delta=0.4$; the full line shows the best Gaussian fit
of width $\Gamma=0.64 \delta$.} 
\label{fig8}
\end{figure}

The function $\rho_{W}$ characterizes the average probability 
distribution of $W_{im}$ (see a numerical example in Fig.3 of \cite{GS}). 
For moderate coupling strength, 
$\rho_{W}$ is well described by the well-known 
Breit-Wigner distribution $\rho_W = \rho_{BW}$:
\begin{equation}
\label{bw}
\rho_{BW}(E-E_{i})= 
\frac{\Gamma}{2 \pi((E-E_{i})^{2}+\Gamma^{2}/4)}
\end{equation}
where $\Gamma$ is the width of the distribution.  This expression is 
valid when $\Gamma$ is smaller than the bandwidth ($\Gamma <
\sqrt{n}\delta$) and many levels are contained inside this width.
In this regime, the Breit-Wigner width $\Gamma$ is given by the Fermi golden
rule: $\Gamma = 2\pi U_s^2 \rho_c$, where $U_s$ is the root mean square
of the transition matrix element and $\rho_c$  is the density of directly
coupled states.  The validity of this formula was well checked in
many-body systems with quantum chaos \cite{zelevinsky,geor1,pichard,guhr}.
In our case $U_s \sim J$ and $\rho_c \sim n/\delta$, so that:
\begin{equation}
\label{gamma}
\Gamma \sim \frac{J^2 n}{\delta}.
\end{equation}

This dependence is confirmed by the data on Fig.8.  However, for large
$J$, when $\Gamma > \sqrt{n}\delta$, the shape of $\rho_W$ becomes
non-Lorentzian and is well fitted by a Gaussian distribution.
The width of this modified distribution grows
like $\Gamma \sim J$.  This scaling naturally appears in the limit
$\delta=0, \;\; J \ll \Delta_0$, since the noninteracting
part of the Hamiltonian is simply a constant commuting with the perturbation.
The change from
one dependence to the other takes place for $J > \delta/ n^{1/4}$.
Above this limit $\Gamma$ is still weakly dependent on
the number of qubits $n$. We expect that for $J \gg \delta$
the energy width of one band is $\Gamma \sim J \sqrt{n}$
(effective frequency of n Rabi frequencies with random 
signs), and have checked numerically this law for $\delta=0$ (data not shown).

\begin{figure}
\epsfxsize=3.0in
\epsfysize=3.0in
\epsffile{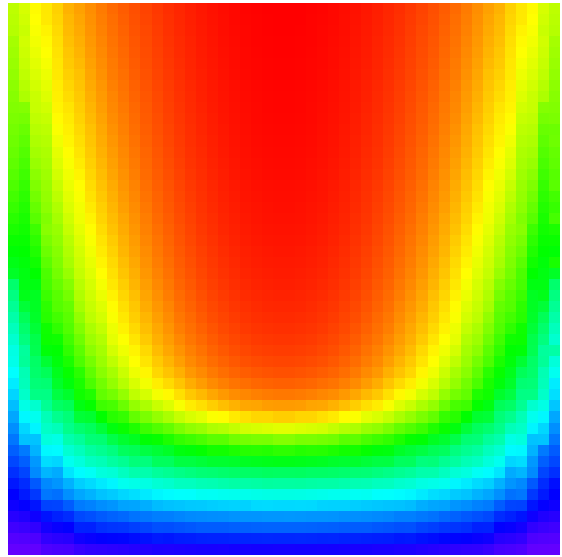}
\caption{Melting of the quantum computer core generated by the inter-qubit
coupling.
Color represents the level of quantum eigenstate entropy $S_q$, from bright red
($S_q \approx 12$) to blue
($S_q =0$).  Horizontal axis is the scaled energy 
$E/\delta$ of the computer eigenstates 
in the central band counted
from the band bottom to the top ($E/\delta \approx \pm \sqrt{n}$).
Vertical axis is the value of $J/\delta$, varying from $0$ to $0.5$. Here
$n=16$, $J_c/\delta=0.22$, and
one random realization is chosen.} 
\label{fig9}
\end{figure}
According to the results obtained from many-body systems \cite{geor1}, 
the number of quantum register states mixed inside the width $\Gamma$
is of the order of $\Gamma \rho_n$, and is exponentially large.  
This however assumes that $J>J_c$ and the system is already in the quantum
chaos regime.  In this case the quantum eigenstate entropy $S_q$ 
is large ($S_q \approx \log_{2} (\Gamma \rho_n) \sim n$) and the
operability of the computer is quickly destroyed, since many quantum register states
become mixed.  The pictorial view of the quantum computer melting is shown on 
Fig.9.  This image is qualitatively similar to the one in \cite{GS} (Fig.5
there), which was obtained for the SGQC model at $\delta=\Delta_0$.  In Fig.9
the melting goes in a smoother way since all the states belong to the same
central band (quantum computer core).

\begin{figure}
\epsfxsize=2.8in
\epsfysize=2.8in
\epsffile{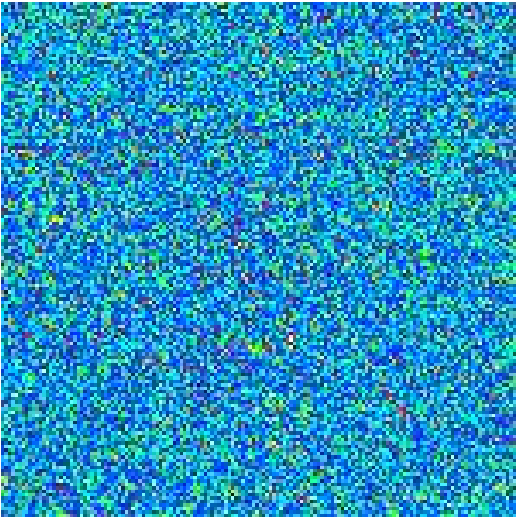}
\caption{Quantum chaos in the quantum register:
color represents the value of the projection 
probability $W_{im}$ of the quantum register states
on the eigenstates of the Hamiltonian,
from bright red (maximal value)to blue
(minimal value).  Horizontal axis corresponds to 150 quantum register states,
the vertical axis represents the 150 computer eigenstates
(both ordered in energy). Here
$n=16$, $J/\delta=0.4$ ($J/\delta > J_c/\delta=0.22$), and
one random realization is chosen.} 
\label{fig10}
\end{figure}

\begin{figure}
\epsfxsize=2.8in
\epsfysize=2.8in
\epsffile{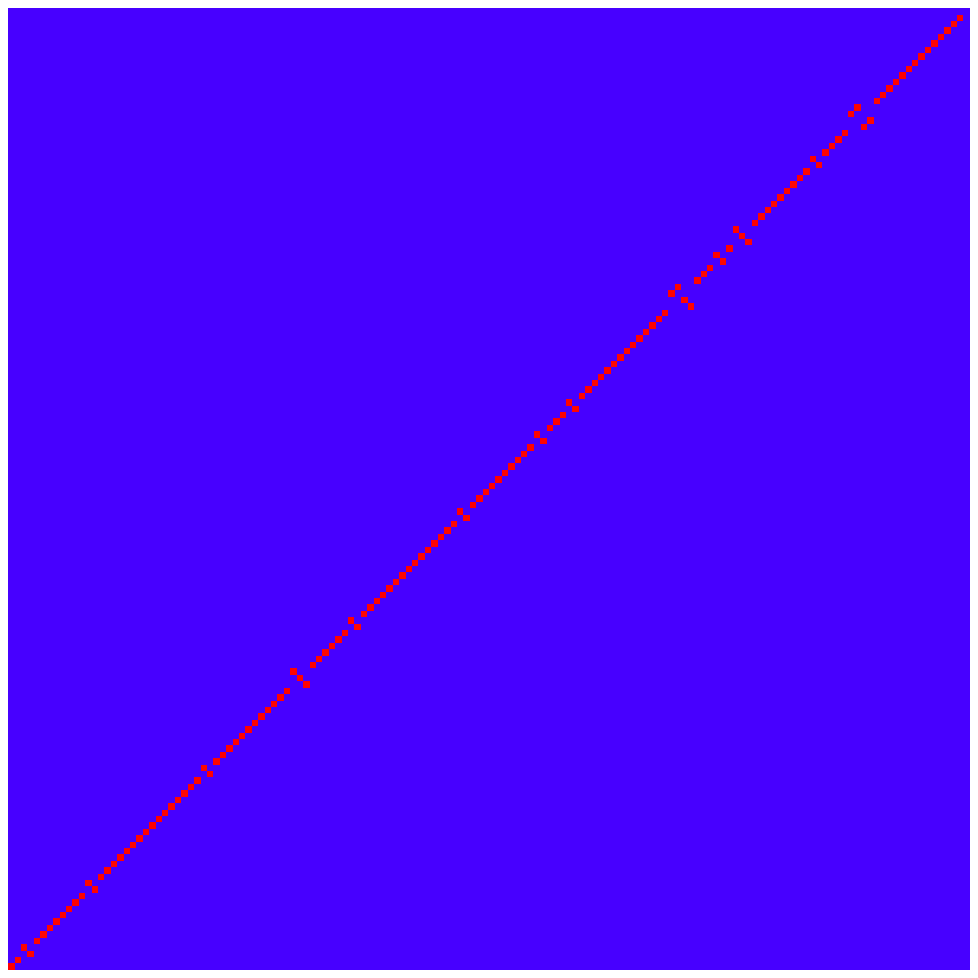}
\caption{Same as Fig.10 below the quantum chaos border,
$J/\delta=0.001$ ($J/\delta \ll J_{cs}/\delta=0.026$).} 
\label{fig11}
\end{figure}

\begin{figure}
\epsfxsize=2.8in
\epsfysize=2.8in
\epsffile{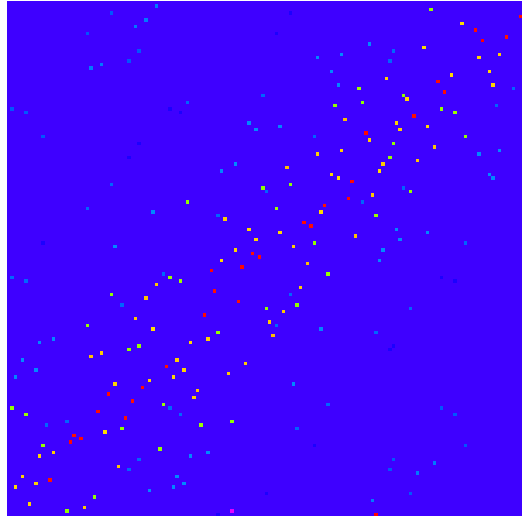}
\caption{Same as Fig.10 for $J/\delta=0.01$ 
($J/\delta \sim J_{cs}/\delta=0.026$)} 
\label{fig12}
\end{figure}

The effect of quantum chaos melting 
in the quantum register representation is shown on Fig.10 for $J > J_c$.
The ideal register structure is manifestly washed out.
On the contrary, below the chaos border ($J<J_c$), 
only few quantum register states are mixed. 
For comparison, Fig.11 shows the same part of the register in the regime
$J \ll J_{cs}$ (no mixing of states) and Fig.12 in the regime 
$J \sim J_{cs}$ 
(few states are mixed).

\section{Time evolution in the SGQC model}

In the previous section we determined the properties of eigenstates
of the quantum computer in the presence of residual inter-qubit coupling.
In the presence of this coupling the quantum register states $|\psi_i>$
are not
any more stationary states, and therefore it is natural to analyze
how they evolve in time.  Indeed, if at time $t=0$ an initial state is 
$|\chi (t=0)>=|\psi_{i_0}>$ corresponding to the quantum register state $i_0$, 
then with time the probability will spread over the register and 
at a time $t$ the projection probability on the register state $|\psi_i>$
will be:
\begin{equation}
\label{proj}
\begin{array}{l}
F_{ii_0} (t) = |<\psi_i|\chi (t)>|^2 \\
= \sum_{m,m'} A_{im} A_{i_0m}^{*} A_{im'}^{*}
A_{i_0m'} \exp (i(E_{m'}-E_m)t),
\end{array}
\end{equation}
where $A_{im}=<\psi_i|\phi_m>$ and $E_m$ is the energy of the stationary state
$|\phi_m>$ and we chose $\hbar=1$.  For $J \ll J_c$, the probability $F_{i_0i_0}(t)$ is very close
to one for all times since the states are not mixed by the interaction. 
This means that all quantum register states $|\psi_i>$ remain well defined,
and the computer can operate properly.  For $J \sim J_{cs}$, 
only few states $|\psi_i>$ are
mixed by the interaction,  
and $F_{i_0i_0}(t)$ oscillates in time regularly around an average value
of order $1/2$.  These oscillations are similar to the Rabi oscillations
between two levels with frequency $\Omega \sim J$. 
An example is presented in Fig.13.  
In this regime,
we expect that error-correcting codes \cite{shor2,steane1} may efficiently
correct the spreading over few quantum register states.
\begin{figure}
\epsfxsize=3.4in
\epsfysize=2.6in
\epsffile{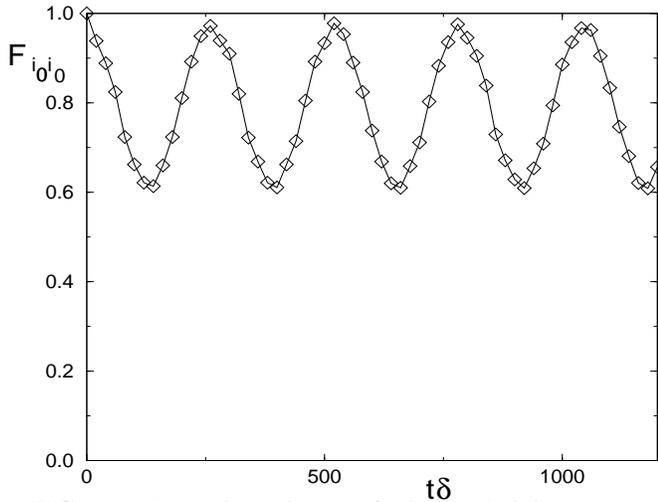}
\caption{Time-dependence of the probability to remain in the same quantum
register state for $n=16$, $J =0.01 \sim J_{cs}=0.026$
( $J_c/\delta=0.22$) and
one random realization is chosen.} 
\label{fig13}
\end{figure}
\begin{figure}
\epsfxsize=3.4in
\epsfysize=2.6in
\epsffile{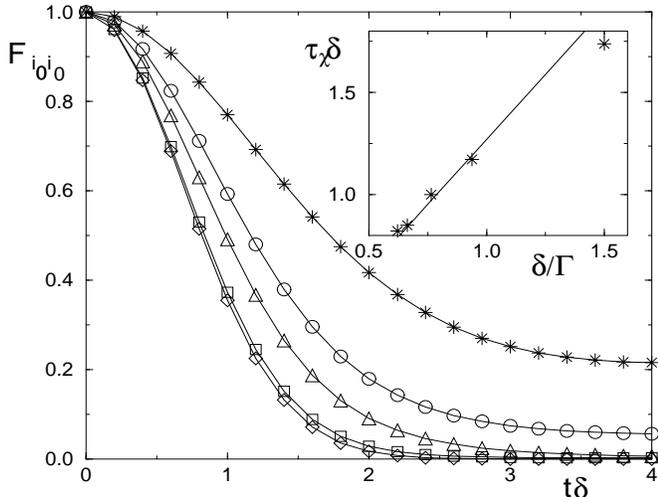}
\caption{Time-dependence of the probability to remain in the same quantum
register state for $J/\delta=0.4 \gg J_{c}/\delta$. Data are shown for
$n=16$ (diamonds, $J_c/\delta=0.22$); $n=15$ (squares, $J_c/\delta=0.24$);
$n=12$ (triangles, $J_c/\delta=0.28$); $n=9$ (circles, $J_c/\delta=0.35$);
$n=6$ (stars, $J_c/\delta=0.59$). Average is made over 200 states randomly
chosen in the central band. Insert shows the chaotic time scale $\tau_{\chi}$ 
(defined by $F_{i_0i_0}(\tau_{\chi})=1/2$) as a
function of $1/\Gamma$; the straight line is $\tau_{\chi}=1.27/\Gamma$.} 
\label{fig14}
\end{figure}

For $J > J_{cs}$, quantum chaos sets in, and with time the probability spreads
over more and more quantum register states until a quasi-stationary regime
is reached where an exponentially large number of states are mixed.  The
probability $F_{i_0i_0}(t)$ drops approximately to zero, as shown on Fig.14.
The chaotic time scale for this decay $\tau_{\chi}$ can be estimated 
as $\tau_{\chi} \sim 1/\Gamma$
where $\Gamma$ is the width determined in the previous section.  This estimate
is very natural in the Fermi golden rule regime, with Breit-Wigner
local density of states (\ref{bw}), since $F_{i_0i_0}(t)$ is essentially
the Fourier transform of the local density of states $\rho_{W}$, and therefore
decreases as $\exp (-\Gamma t)$.
We note that the decay in this regime was recently discussed in
\cite{flambaum}.
According to our data, when $\Gamma$ becomes comparable to the energy bandwidth 
$\sqrt{n}\delta$, $\rho_W$ is close to a Gaussian distribution of width 
$\Gamma$, and its Fourier transform $F_{i_0i_0}(t)$ is also a Gaussian 
of width $1/\Gamma$.  Therefore in both regimes we expect the time scale $\tau_{\chi}$
for the decay of $F_{i_0i_0}(t)$ to be $\tau_{\chi} \sim 1/\Gamma$.  The data
shown on Fig.14 correspond to the saturation regime for large values of $n$,
and the insert shows that $\tau_{\chi} \sim 1/\Gamma$ is still valid.
In fact the curve for $n=16$ in Fig.14 is already close
to the limiting decay curve at $\delta=0$ (data not shown).

\begin{figure}
\epsfxsize=3.4in
\epsfysize=2.6in
\epsffile{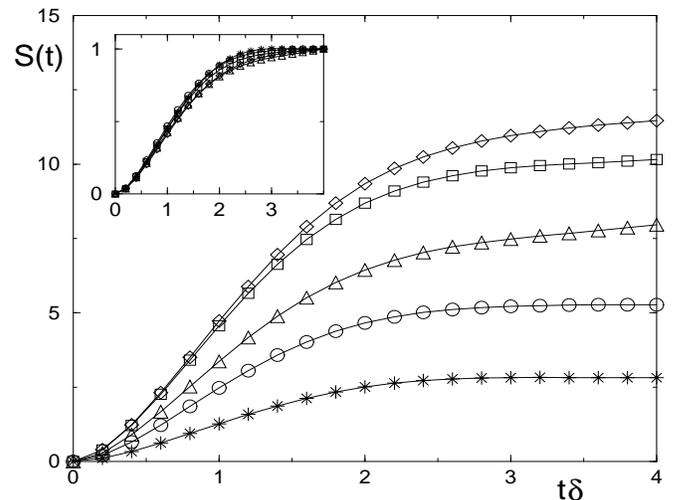}
\caption{Time-dependence of the quantum entropy $S(t)$ for
 $J/\delta=0.4 \gg J_{c}/\delta$; symbols are as in Fig.14.
Average is made over 200 initial states randomly
chosen in the central band. Insert shows the same curves normalized to
their maximal value.} 
\label{fig15}
\end{figure}

At the same time scale $\tau_{\chi}$ the quantum  entropy $S(t)$ is large but 
is still growing. It reaches its maximal value on a larger time scale
which seems independent of $n$.  At this stage, 
 an initial quantum register state is now
spread over most of the register 
(Here $S(t) = -\sum_{i} F_{ii_0}(t) \log_2 F_{ii_0}(t)$).  
This process is shown on Fig.15.  
This maximal value of $S(t)$ is approximately given by $S_q$ (see Fig.6)
and accordingly decreases with decreasing $J$ as is illustrated in Fig.16.

Fig.17 illustrates this mixing process in the quantum register representation,
evolving in time.  The quantum computer hardware becomes quickly destroyed
due to the inter-qubit coupling.
It is necessary to decrease the coupling strength below the quantum chaos
border to get well-defined quantum register states for $t>0$, as is illustrated
in Fig.18.

\begin{figure}
\epsfxsize=3.4in
\epsfysize=2.6in
\epsffile{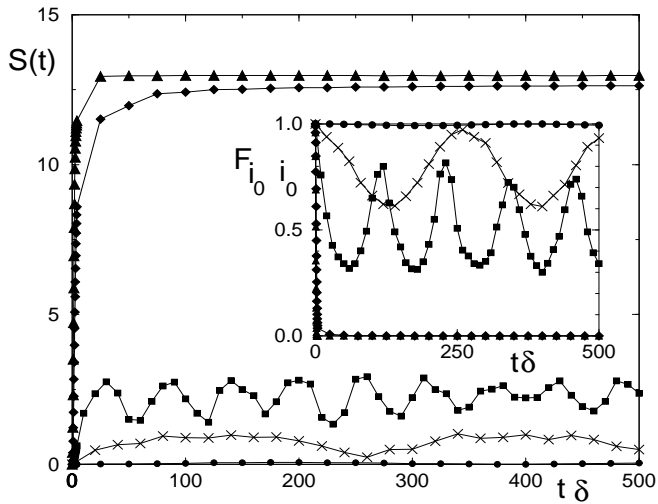}
\vglue 0.3cm
\caption{Time-dependence of the quantum entropy $S(t)$ for 
different values of $J$,
$n=16$ $J_c/\delta=0.22$, $J_{cs}/\delta=0.026$, and
one random realization is chosen: 
$J/\delta=0.001 \ll J_{cs}/\delta$ (disks);
$J/\delta=0.01 < J_{cs}/\delta$ (crosses); 
$J/\delta=0.03 \approx J_{cs}/\delta$ (squares);
$ J/\delta=0.2 \approx J_{c}/\delta$ (diamonds); 
$J/\delta=0.4 > J_{c}/\delta$ (triangles).
Insert gives the probability to remain in the same quantum
register state for the same values of $J/\delta$.
Averages are made over 200 states randomly
chosen in the central band.} 
\label{fig16}
\end{figure}

\vskip 0.5cm

\begin{figure}
\epsfxsize=3.0in
\epsfysize=3.0in
\epsffile{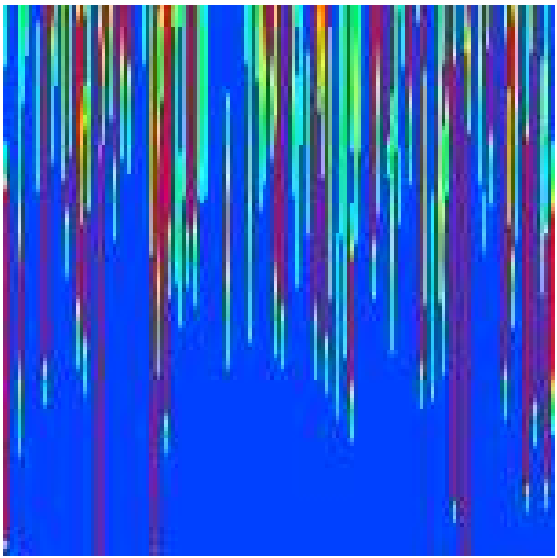}
\vglue 0.3cm
\caption{Time explosion of quantum chaos in the quantum register:
color represents the value of the projection probability 
$|<\psi_i|\chi(t)>|^2$ of an initial state
on the quantum register states ordered in energy,
from bright red (maximal value)to blue
(minimal value).  Horizontal axis corresponds to 150 states,
the vertical axis to 150 time steps, from $t\delta=0$ to
$t\delta=2$. At $t\delta=0$, the chosen initial state is the
superposition of two quantum register states. Here
$n=16$, $J/\delta=0.4$ ($J/\delta > J_c/\delta=0.22$), and
one random realization is chosen.} 
\label{fig17}
\end{figure}

\begin{figure}
\epsfxsize=3.0in
\epsfysize=3.0in
\epsffile{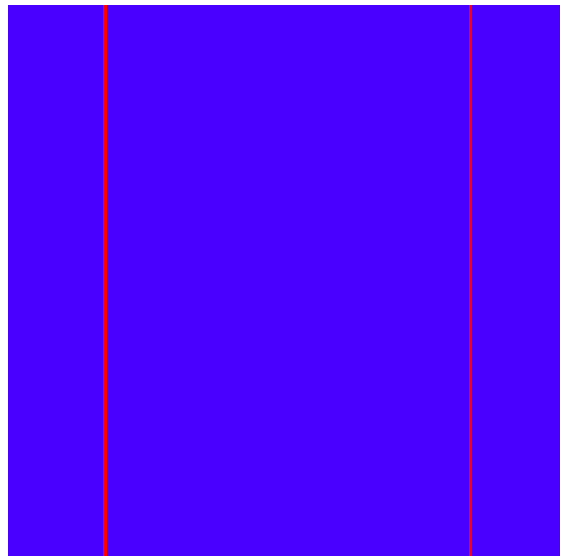}
\vglue 0.3cm
\caption{Same as Fig.17 below the quantum chaos border, 
 $J/\delta=0.001$ ($J/\delta \ll J_{cs}/\delta=0.026$).} 
\label{fig18}
\end{figure}

The obtained data clearly show that exponentially many quantum register states
become mixed after the finite chaotic time scale $\tau_{\chi} \approx 1/\Gamma$.

\section{conclusions}
The results presented in this paper show that residual inter-qubit coupling
can lead to quantum chaos and ergodic very complicated eigenstates of the
quantum computer.  We have shown that in this regime the quantum register
states disintegrate quickly in time over an exponentially large number of
states and the computer operability is destroyed.  
We determined the dependence of the chaotic time scale 
$\tau_{\chi}$ of this process on
coupling strength $J$, detuning fluctuations $\delta$ of one-qubit 
energy spacing, and number of qubits $n$.  
After this time $\tau_{\chi}$ the quantum computer hardware is melt.
To prevent this melting one needs to introduce an efficient
error-correcting code which operates on a time scale much shorter
than $\tau_{\chi}$ and suppresses the development of quantum chaos.
To avoid the quantum chaos regime dangerous 
for quantum computing, one should
engineer the quantum computer in the integrable regime below the quantum
chaos border $J_c \approx 3 \delta/n$.  It is important to note that this
border decreases with the detuning $\delta$, showing that imperfections
do not all conspire against the operability of the computer.  We stress
again that the transition to quantum chaos is an internal process which
happens in a perfectly isolated system
with no coupling to external world.
Nevertheless, since a decoherence can
be viewed as a result of internal interactions in a larger system, 
the results presented
here may also apply to this problem.

Our main conclusion is that although in the quantum chaos regime a quantum
computer cannot operate for long, 
fortunately the border for this process happens to be exponentially
larger than the spacing between adjacent computer eigenstates, and therefore a 
broad parameter region remains available for realization of a quantum computer.
Another possibility is to operate the quantum computer in the regime of
quantum chaos. However, here one should keep in mind that 
after the chaotic time scale $\tau_{\chi}$ the computer hardware
will melt due to inter-qubit coupling
and quantum chaos. Therefor, the computer operability in this regime
is possible only if many gate operations can be realized 
during the finite time $\tau_{\chi}$ (in a sense it becomes
similar to the decoherence time). It is clear that
the most preferable regime corresponds to quantum computer
operation below the quantum chaos border.

We thank O.P. Sushkov for stimulating
discussions, and the IDRIS in Orsay and the CICT in Toulouse for access to 
their supercomputers. One of us (DLS) thanks the Gordon
Godfrey foundation at the University of New South Wales at Sydney
for the hospitality at the final stage of this work.
This research is partially done in the frame of EC program RTN1-1999-00400.


\vskip -0.5cm
\begin{thebibliography}{99}
\bibitem[*]{byline1} http://w3-phystheo.ups-tlse.fr/$\sim$dima
\bibitem{feynman} R.~P.~Feynman,
   Found. Phys. {\bf 16}, 507 (1986).
\bibitem{DiVi1} D.~P.~Di~Vincenzo, Science {\bf 270}, 255 (1995).
\bibitem{josza} A.~Ekert and R.~Josza, Rev. of Mod. Phys. {\bf 68}, 733 (1996).
\bibitem{steane2}  A. Steane, Rep. Progr. Phys. {\bf 61},
117 (1998).
\bibitem{deutsch} D.~Deutsch, Proc. R. Soc. London Ser. A {\bf 425}, 73 (1989).
\bibitem{DiVi2} D.~P.~Di~Vincenzo, Phys. Rev. A {\bf 51}, 1015 (1995).
\bibitem{shor1} P.~W.~Shor, in Proc. 35th Annu. Symp. Foundations of
Computer Science (ed. Goldwasser, S. ), 124 (IEEE Computer Society, Los
Alamitos, CA, 1994).
\bibitem{Grover} L.~K.~Grover, Phys. Rev. Lett. {\bf 79}, 325 (1997).
\bibitem{shor2} A.~R.~Calderbank  and P.~W.~Shor, 
Phys. Rev. A {\bf 54}, 1098 (1996).
\bibitem{steane1} A.~Steane,
Proc. Roy. Soc. Lond. A {\bf 452}, 2551 (1996).
\bibitem{zoller} J.~I.~Cirac and P.~Zoller,
Phys. Rev. Lett. {\bf 74}, 4091 (1995).
\bibitem{nmr} N.~A.~Gershenfeld and I.~L.~Chuang, 
 Science {\bf 275}, 350 (1997); D.~G.~Cory, A.~F.~Fahmy and T.~F.~Havel,
In Proc. of the 4th Workshop on Physics and Computation
(Complex Systems Institute, Boston, MA, 1996).
\bibitem{vagner} V.~Privman, I.~D.~Vagner and G.~Kventsel, 
Phys. Lett. A {\bf 239}, 141 (1998).
\bibitem{kane} B.~E.~Kane, Nature {\bf 393}, 133 (1998).
\bibitem{bowden} C.~D.~Bowden and S.~D.~Pethel, Int. J. of Laser Phys., to
appear (2000), (quant-ph/9912003).
\bibitem{loss} D.~Loss  and D.~P.~Di~Vincenzo,
 Phys. Rev. A {\bf 57}, 120 (1998).
\bibitem{cooper} Y.~Nakamura, Yu.~A.~Pashkin, and J.~S.~Tsai,  
     Nature {\bf 398}, 786 (1999).
\bibitem{lattice} G.~K.~Brennen, C.~M.~Caves, P.~S.~Jessen and I.~H.~Deutsch
     Phys. Rev. Lett. {\bf 82}, 1060 (1999); D.~Jaksch, H.~J.~Briegel, 
     J.~I.~Cirac, C.~W.~Gardiner and P.~Zoller,
     Phys. Rev. Lett. {\bf 82}, 1975 (1999).
\bibitem{helium} P.~M.~Platzman and M.~I.~Dykman, Science {\bf 284}, 1967
    (1999).
\bibitem{monroe} C.~Monroe, D.~M.~Meekhof, B.~E.~King, W.~M.~Itano
    and  D.~J.~Wineland, Phys. Rev. Lett. {\bf 75}, 4714 (1995).
\bibitem{3q} L.~M.~K.~Vandersypen, M.~Steffen, M.~H.~Sherwood,
    C.~S.~Yannoni, G.~Breyta and I.~L.~Chuang, quant-ph 9910075.
\bibitem{haroche} S.~Haroche and J.~M.~Raimond, Phys. Today. 51 (august 1996).
\bibitem{vagner2} D.~Mozyrsky, V.~Privman and I.~D.~Vagner,
    cond-mat/0002350.
\bibitem{paz} C.~Miquel, J.~P.~Paz and R.~Perazzo, Phys. Rev. A {\bf 54},
    2605 (1996).
\bibitem{zurek} C.~Miquel, J.~P.~Paz and W.~H.~Zurek, 
    Phys. Rev. Lett. {\bf 78},
    3971 (1997).
\bibitem{french} J.~B.~French and S.~S.~M.~Wong, 
   Phys. Lett. B {\bf 33}, 449 (1970); O.~Bohigas and J.~Flores, 
   {\em ibid} {\bf 34}, 261 (1971).
\bibitem{aberg} S. {\AA}berg, Phys. Rev. Lett. {\bf 64}, 3119 (1990).
\bibitem{zelevinsky} V.~Zelevinsky, B.~A.~Brown, N.~Frazier and M.~Horoi, 
   Phys. Rep. {\bf 276}, 85 (1996); V.~V.~Flambaum, F.~M. Izrailev, and G.~Casati, 
  Phys. Rev. E {\bf 54}, 2136 (1996); V.~V.~Flambaum and F.~M. Izrailev, 
  Phys. Rev. E {\bf 56}, 5144 (1997).
\bibitem{sivan} U.~Sivan,  F.~P.~Milliken, K.~Milkove, S.~Rishton,
  Y.~Lee, J.~M.~Hong, V.~Boegli, D.~Kern, and M.~de~Franza,
  Europhys. Lett. {\bf 25}, 605 (1994).
\bibitem{1997} D.~L.~Shepelyansky and O.~P.~Sushkov,
  Europhys. Lett. {\bf 37}, 121 (1997).
\bibitem{jacquod} P.~Jacquod and  D.~L.~Shepelyansky, 
  Phys. Rev. Lett. {\bf 79}, 1837 (1997).
\bibitem{mirlin} A.~D.~Mirlin and Y.~V.~Fyodorov,
  Phys. Rev. B {\bf 56}, 13393 (1997).
\bibitem{geor1} B.~Georgeot and D.~L.~Shepelyansky,
  Phys. Rev. Lett. {\bf 79},
  4365 (1997).
\bibitem{pichard} D.~Weinmann, J.-L.~Pichard and Y.~Imry,
  J. Phys. I France
  {\bf 7}, 1559 (1997).
\bibitem{georgeot} B.~Georgeot and D.~L.~Shepelyansky,
  Phys. Rev. Lett. {\bf 81},
  5129 (1998).
\bibitem{houches} {\it Les Houches Lecture Series} {\bf 52},
        Eds.  M.-J.~Giannoni, A.~Voros and J.~Zinn-Justin (North-Holland,
        Amsterdam, 1991).
\bibitem{guhr} T. Guhr, A. M\"uller-Groeling and H.~A.~Weidenm\"uller, 
    Phys. Rep. {\bf 299}, 189 (1999).
\bibitem{GS} B.~Georgeot and D.~L.~Shepelyansky, quant-ph/9909074.
\bibitem{molmer} A.~S\o rensen and K.~M\o lmer,
Phys. Rev. Lett. {\bf 83}, 2274 (1999).
\bibitem{longrange} This result is valid for short-range couplings.  In the case
of long-range couplings where all qubits are coupled, the system is similar to the
one studied in \cite{georgeot}, $\Delta_c \sim \delta/n^2$ and we will have
$J_c \approx C'\delta/n^2$, where $C'$ is a numerical constant.
\bibitem{note}  For even $n$ and $\delta=0$ the system has an additional
 symmetry corresponding to the inversion of all $\bf{\sigma_i} \rightarrow 
 -\bf{\sigma_i}$ (spin inversion).  However,
 inside one symmetry class
 the spacing statistics is still close to $P_W(s)$ 
 according to our numerical data (not shown).
\bibitem{wigner} E.~P.~Wigner, Ann. Math. {\bf 62}, 548 (1955); {\bf 65},
203 (1957).
\bibitem{flambaum} V.~V.~Flambaum, quant-ph/9911061.
\end{thebibliography}
\end{document}